# Two-probe study of hot carriers in reduced graphene oxide

Brian A. Ruzicka[1], Nardeep Kumar[1], Shuai Wang[2], Kian Ping Loh[2], and Hui Zhao[1, a)]
[1]*Department of Physics and Astronomy, The University of Kansas, Lawrence, Kansas 66045, USA*
[2]*Department of Chemistry, National University of Singapore, 3 Science Drive 3, Singapore 117543*

The energy relaxation of carriers in reduced graphene oxide thin films is studied using optical pump-probe spectroscopy with two probes of different colors. We measure the time difference between peaks of the carrier density at each probing energy by measuring a time-resolved differential transmission and find that the carrier density at the lower probing energy peaks later than that at the higher probing energy. Also, we find that the peak time for the lower probing energy shifts from about 92 to 37 fs after the higher probing energy peak as the carrier density is increased from $1.5 \times 10^{12}$ to $3 \times 10^{13}/cm^2$, while no noticeable shift is observed in that for the higher probing energy. Assuming the carriers rapidly thermalize after excitation, this indicates that the optical phonon emission time decreases from about 50 to about 20 fs and the energy relaxation rate increases from 4 to 10 meV/fs. The observed density dependence is inconsistent with the phonon bottleneck effect.

Graphene, a single layer of carbon atoms, is a very attractive material for many applications including transistors,[1,2] solar cells,[3] electromechanical resonators,[4] ultracapacitors,[5] and composite materials.[6] Charge carriers play a central role in most of these applications. In particular, it has been shown that the mean-free path of carriers in graphene is several hundred nanometers even at room temperature.[1,7] Hence, even in micrometer-sized devices, carriers injected with a high kinetic energy only undergo few or even no phonon scattering events during the transport through the device, maintaining a temperature much higher than the lattice temperature. Since the device is dominated by the hot carriers, it is important to understand and control the dynamics of hot carriers in graphene.

Over the past two years, significant progress has been made on using ultrafast laser techniques to study hot carriers in graphene.[8–19] Most studies so far have focused on measuring the energy relaxation of the carriers by fitting the decay of a differential transmission signal. This tool can be very valuable as it can give much insight into the time scales and mechanisms behind the relaxation of carriers, but it also has its limitations. First, the decay time is often comparable to the temporal width of the probe pulse. Second, a slow decay component related to carrier recombination is often seen, making the decay multiple exponential. Both of these factors limit the accuracy of such measurements in determining the energy relaxation time of excited carriers. Furthermore, previous studies have mainly been limited to graphene samples fabricated by thermal reduction of silicon carbide substrates,[8–16] mechanically exfoliated graphene on Si/SiO$_2$ substrates[17] and graphene thin films grown by chemical vapor deposition.[14] While studies of reduced graphene oxide are relatively rare,[16,18,19] it can be argued that reduced graphene oxide may be one of the most promising type of graphene for use in industry, as it can be produced at low cost[20] and is favorable for large-scale production of graphene-based electronics.[21]

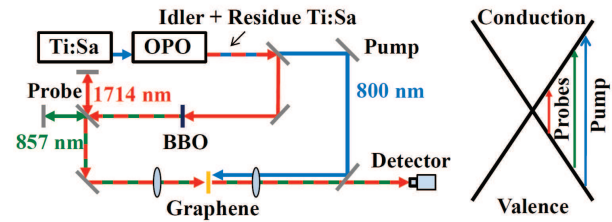

FIG. 1. Experimental setup (left panel) and pump/probe scheme (right panel).

Here we report a study of hot carriers in reduced graphene oxide thin films using ultrafast optical pump-probe spectroscopy with two probes of different colors. First, carriers are injected by a pump pulse. By precisely overlapping the two probes in time, and measuring the time between which the differential transmission signals peak with each probe, we are able to monitor the density of carriers at two different energies for various times after excitation. From these measurements, we observe that the peak carrier density at the lower energy probe occurs about 92 fs later than the higher energy probe, and this decreases to about 37 fs, as the carrier density is increased from $1.5 \times 10^{12}$ to $3 \times 10^{13}/cm^2$. Under the assumption that the carrier thermalization is much faster than the time scales of the study, we can deduce energy relaxation rates on the order of several meV/fs, and optical phonon emission times on the order of several tens of fs. Furthermore, the observed increase in energy relaxation rate with increasing carrier density is in opposition to what one would expect from a phonon bottleneck effect.

The reduced graphene oxide samples are fabricated by spin coating graphene oxide flakes on quartz substrates. The formed films are then transformed to graphene films by thermal reduction at 1000°C.[21] The number of graphene layers is determined to be about 50 by using an

a)Electronic mail: huizhao@ku.edu

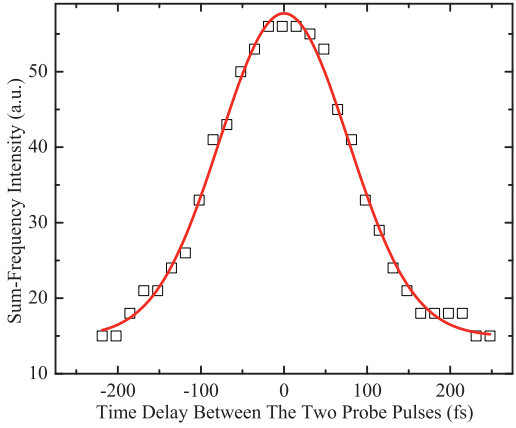

FIG. 2. Cross correlation measurement of the 1714- and 857-nm probe pulses using sum-frequency generation in a [110] GaAs crystal.

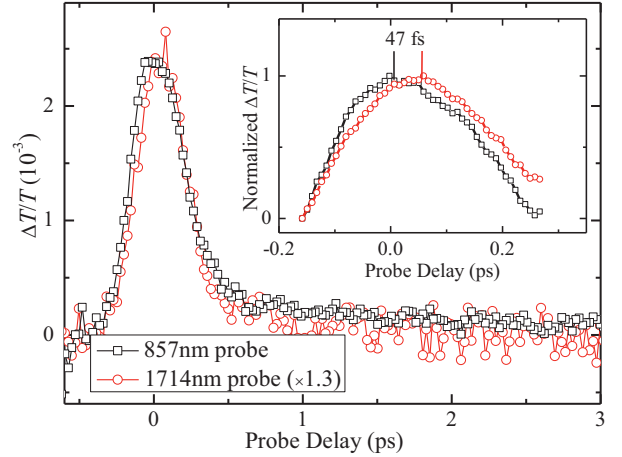

FIG. 3. The differential transmission for both the 857-nm (squares) and the 1714-nm (circles) probes with an average areal carrier density of $2.3 \times 10^{13}/\text{cm}^2$. A slight displacement between the two peaks can be seen. The inset shows another scan around the peaks, and from this we are able to determine that the differential transmission measured with the 1714-nm probe peaks approximately 47 fs later than the 857-nm probe.

atomic force microscope. The absorption of the samples at 800 nm is measured to be about 50%.

Figure 1 summarizes the experimental setup (left panel) and pump/probe scheme (right panel). Carriers are first excited with an 800-nm, 100-fs pump pulse, which is focused to a size of approximately 2.3 $\mu$m at full width at half maximum (FWHM). The pump pulse is obtained from a Ti:sapphire laser (Ti:Sa) with a repetition rate of about 80 MHz. To detect the carriers, we use one of two different probe pulses of central wavelengths 1714 and 857 nm, respectively. The 1714-nm probe is produced from the idler output of an optical parametric oscillator (OPO), which is pumped by the Ti:Sa, and the 857-nm probe is obtained from second-harmonic generation of the 1714-nm probe using a beta barium borate (BBO) crystal. The pulse width of each is 80 fs and 150 fs, respectively, and both are focused to a spot size of approximately 1.5 $\mu$m (FWHM). The differential transmission of the probe pulse, $\Delta T/T_0 \equiv [T(n) - T_0]/T_0$, i.e. the normalized difference in transmission with [$T(n)$] and without ($T_0$) carriers, is measured by modulating the intensity of the pump pulse with a mechanical chopper with a frequency of about 2 kHz and using a lock-in amplifier.

For our experiment, it is crucial to precisely control the time of each probe pulse so that the two probes arrive at the sample at the same time. To do that, we use sum-frequency generation in a GaAs sample grown along [110] direction, which is mounted directly next to the graphene sample. The squares in Fig. 2 show the intensity of the sum-frequency signal with a central wavelength of 571 nm as a function of the time delay between the two probe pulses. From a Gaussian fit to the data (solid line), we determine the FWHM of this cross-correlation is about 180 fs. By repeatedly obtaining the maximum sum-frequency generation, we conclude that we are able to consistently overlap the two probe pulses in time with an error smaller than 5 fs.

We measure the differential transmission with each probe as a function of time by changing the delay between the probe and pump pulses with an average areal carrier density of $2.3 \times 10^{13}/\text{cm}^2$, as shown in Fig. 3. The 0-ps probe delay is defined arbitrarily here, as we are only interested in the difference in time between the peaks of the two curves. In fact, the 0-ps probe delay is expected to be very close to the peak of the 857-nm probe since the photon energies of the pump and the 857-nm probe are close. The inset of Fig. 3 shows another scan around the peaks. The peak of the differential transmission with the 1714-nm probe occurs approximately 47 fs after the peak of the 857-nm probe.

One of the difficulties in studies of hot carrier dynamics in graphene is that the dynamics are often too short to resolve. When the decay time of the signal is comparable to the temporal width of the pulses used, the result can be severely influenced by the convolution of the pulses. In our two-color scheme, however, we can accurately determine the time delay between the two peaks even though it is shorter than the pulses: The convolution of the finite probe pulse with the actual signal only shifts the time of the peak of both probes; the relative time delay between the two peak is not changed. Therefore, it is sufficient to consider only the convoluted pump-probe measurements when determining the relative position of the two peaks.

In order to investigate the dependence of the relative peak times on the carrier density, we repeat the measurement with various carrier densities by changing the energy fluence of the pump pulse. The results are summarized in the left panel of Fig. 4. We see a systematical change of the peak time of the 1714-nm probe with the carrier density, while the peak time of the 857-nm probe remains unchanged. At a density of $1.5 \times 10^{12}/\text{cm}^2$, the 1714-nm probe peak occurs about 92 fs later than the 857-nm probe peak. When the density is increased to $3 \times 10^{13}/cm^2$, the 1714-nm peak shifts earlier, occurring about 37 fs after the 857-nm peak.

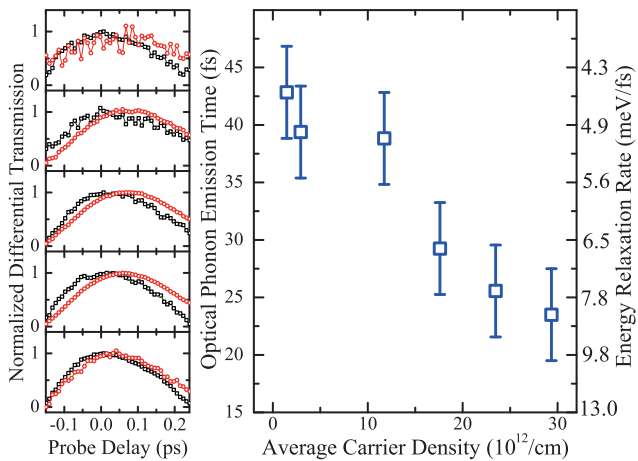

FIG. 4. Left panel: The normalized differential transmission measured with the 857-nm (squares) and the 1714-nm (circles) probes with average areal carrier densities of (from top to bottom) 1.5, 2.9, 11.7, 17.6, and $30 \times 10^{12}/cm^2$, respectively. Right panel: Deduced energy relaxation rate (right axis) and optical phonon emission time (left axis) of hot carriers.

It has been shown by previous ultrafast pump-probe[9,14,15] and nonlinear photoluminescence experiments[22–24] that the thermalization of high density carriers in graphene is extremely fast due to the enhanced carrier-carrier scattering rate by the unique linear energy dispersion. Recent studies have shown that the carrier-carrier scattering time can be as short as 10 fs.[25] Therefore, it is reasonable to assume that the carriers injected by the pump pulse rapidly reach a Fermi-Dirac distribution, with a temperature determined by the pump photon energy. Hence, we present the following possible interpretation of the data: Due to state filling effects, the differential transmission of each probe pulse is proportional to the density of carriers at each probing energy. So, when the differential transmission (and therefore the density of carriers seen by the probe) peaks for the 857-nm (1714-nm) probe, the average energy of the carriers is approximately equal to the central probing energy of 0.72 eV (0.36 eV), which is half of the probe photon energy (right panel of Fig. 1). The difference in the peak times gives the time it takes for the average carrier energy to decrease by 0.36 eV (from 0.72 to 0.36 eV). The dynamics of the holes in the valence band is equally probed, and is expected to be the same as the dynamics of the electrons in the conduction band since the two bands have the same dispersion.

For example, we see from the measurement in Fig. 3 that the energy relaxation of 0.36 eV takes 47 fs. This corresponds to an energy relaxation rate of about 8 meV/fs. We can extend this analysis even further, since it is known that the dominant energy relaxation channel is the emission of G-mode optical phonons with an energy 0.195 eV.[26] Therefore, under the previous assumptions, we deduce that at this carrier density, the optical phonon emission time is $47 \times (0.195/0.36) = 25$ fs.

We can also perform this analysis on our study of the power dependence of the relative peak times. These results are summarized in the right panel of Fig. 4. At a density of $1.5 \times 10^{12}/cm^2$, the optical phonon emission time is about 50 fs, corresponding to an energy relaxation rate of about 4 meV/fs. When the density is increased to $3 \times 10^{13}/cm^2$, the optical phonon emission time decreases to about 20 fs, corresponding to an energy relaxation rate of about 10 meV/fs. We note that the values of the optical phonon emission time as well as the density dependence determined in this way are reasonably consistent with recently theoretical calculations.[27,28]

To examine the validity of such an analysis, we consider three different situations after excitation. The first is the one we have just discussed – after excitation, the electrons rapidly thermalize, then cool on a much slower time scale. In this case we are able to obtain our conclusions about the optical phonon emission time and energy relaxation rate. The next situation is one in which the thermalization is slow, and the energy relaxation is even slower, so that during the duration of the differential transmission signals, we are observing only the thermalization process. This situation can immediately be eliminated as a possibility, due to the fact that we are probing at energies below our excitation energy, and therefore both probes will see the carrier density continually increase as the carriers equilibrate to attain a Fermi-Dirac distribution. We verified this by simulation using the Boltzmann equation, starting with a gaussian distribution of carriers and ending with a Fermi-Dirac distribution. The last situation is one in which both thermalization and energy relaxation occur at comparable rates, i.e. both occur during the duration of the differential transmission signals. In this case, we are not able to determine the average energy of the carriers when the density of carriers at each probe reaches its peak value. However, since we would not see any change in the relative peak times of the carrier density at each probing energy without energy relaxation, we are still obtaining some measure of the energy relaxation. So, while this situation may cause the magnitude of the deduced optical phonon emission times to change, we expect the overall behavior of the optical phonon emission time with carrier density to remain the same.

Finally, it is interesting to note that our experimental results are in opposition to the phonon bottleneck effect. Recently, a phonon bottleneck effect on the energy relaxation of hot carriers in graphene samples fabricated by thermal reduction of silicon carbide has been proposed.[14,15] In these studies, the decay time of the differential transmission, on the order of 1 ps, was found to *increase* with the carrier density. Under the assumption that the carrier recombination time is much longer than the measured decay time, the decay time was attributed to the energy relaxation time. The increase of the energy relaxation time with the carrier density was attributed to the effect of a nonequilibrium distribution of optical phonons emitted by the hot carriers. However, more recent experiments indicated that the carrier recombination time can be as short as sub-ps.[22–24] Our experimental results are in opposition to the phonon bottleneck effect and can be well explained by the density-dependent optical phonon emission time. Furthermore, the energy relaxation times we measured by the two-probe technique are significantly shorter than those deduced from these studies. However, it is possible that the phonon dynamics in the reduced graphene oxides are different from the graphene samples fabricated from silicon carbide. Our results suggest the more studies are needed on this interesting effect.

We have studied photoexcited carriers in reduced graphene oxide samples using a two-color probe technique. By overlapping the two probes in time, and measuring the time at which the differential transmission signal peaks for each probe, we are able to directly measure the time at which the carrier density peaks at the two probing energies. Under the assumption that the carriers rapidly thermalize after excitation, this can be interpreted as an average carrier energy decrease from 0.72 to 0.36 eV in approximately 47 fs with a carrier density of $2.3 \times 10^{13}/cm^2$. This corresponds to an energy relaxation rate of about 8 meV/fs. Since the energy relaxation of carriers is mainly caused by the emission of 195 meV G-mode optical phonons, we deduce an optical phonon emission time of about 25 fs. Furthermore, we found that the optical phonon emission time decreases from about 50 to about 20 fs, and the energy relaxation rate increases from 4 to 10 meV/fs, as the carrier density is increased from $1.5 \times 10^{12}$ to $3 \times 10^{13}/cm^2$. The observed density dependence in our graphene samples is inconsistent with the phonon bottleneck effect that was observed in graphene samples fabricated on silicon carbide substrates.

We thank Wang-Kong Tse for useful discussions on optical phonon scattering in graphene. We acknowledge support from the US National Science Foundation under Awards No. DMR-0954486 and No. EPS-0903806, and matching support from the State of Kansas through Kansas Technology Enterprise Corporation. We thank the support of NRF-CRP "Graphene Related Materials and Devices" (Grant No. R-143-000-360-281). Acknowledgment is also made to the Donors of the American Chemical Society Petroleum Research Fund for support of this research.